\documentclass[preprint]{emulateapj}
\usepackage{graphicx}
\usepackage{amssymb}
\usepackage{amsmath}
\usepackage{natbib}
\usepackage{url}
\nocite{*}

\submitted{Accepted to ApJL, May 30, 2017}

\shortauthors{Tofflemire et al.}
\shorttitle{Pulsed Accretion in TWA 3A}
\begin{document}

\title{Pulsed Accretion in the T Tauri Binary TWA 3A}

\author{Benjamin M.\ Tofflemire\altaffilmark{1},
Robert D.\ Mathieu\altaffilmark{1},
Gregory J.\ Herczeg\altaffilmark{2},
Rachel L.\ Akeson\altaffilmark{3} \&
David R.\ Ciardi\altaffilmark{3}}

\altaffiltext{1}{Department of Astronomy, University of Wisconsin--Madison,
  475 North Charter Street, Madison, WI 53706, USA}
\altaffiltext{2}{The Kavli Institute for Astronomy and Astrophysics, Peking
  University, Beijing 100871, China} 
\altaffiltext{3}{NASA Exoplanet Science Institute, IPAC/Caltech, Pasadena, CA
  91125, USA}

\begin{abstract}

  TWA 3A is the most recent addition to a small group of young binary systems
  that both actively accrete from a circumbinary disk and have spectroscopic
  orbital solutions. As such, it provides a unique opportunity to test binary
  accretion theory in a well-constrained setting. To examine TWA 3A's
  time-variable accretion behavior, we have conducted a two-year, optical
  photometric monitoring campaign, obtaining dense orbital phase coverage
  ($\sim$20 observations per orbit) for $\sim$15 orbital periods. From
  $U$-band measurements we derive the time-dependent binary mass accretion
  rate, finding bursts of accretion near each periastron passage. On average,
  these enhanced accretion events evolve over orbital phases 0.85 to 1.05,
  reaching their peak at periastron. The specific accretion rate increases
  above the quiescent value by a factor of $\sim$4 on average but the peak can
  be as high as an order of magnitude in a given orbit. The phase dependence
  and amplitude of TWA 3A accretion is in good agreement with numerical
  simulations of binary accretion with similar orbital parameters. In these
  simulations, periastron accretion bursts are fueled by periodic streams of
  material from the circumbinary disk that are driven by the binary orbit. We
  find that TWA 3A's average accretion behavior is remarkably similar to DQ
  Tau, another T Tauri binary with similar orbital parameters, but with
  significantly less variability from orbit to orbit. This is only the second
  clear case of orbital-phase-dependent accretion in a T Tauri binary.

\end{abstract}


\section{INTRODUCTION}
\label{intro}

TWA 3, also known as Hen 3-600, is a pre-main-sequence (pre-MS) star system
composed of two spatially resolved components: TWA 3A, a spectroscopic binary
hosting a circumbinary accretion disk, and TWA 3B, a diskless tertiary at a
separation of 1$\farcs$5 ($\sim$52.5 AU in projection;
\citealt{delaRezaetal1989,Jayawardhanaetal1999,Andrewsetal2010}; Kellogg et
al.\ 2017). This multi-star architecture offers a unique
opportunity to investigate the impact multiplicity has on the distribution and
evolution of circumstellar material during star formation. In this Letter we
focus on the spectroscopic binary, TWA 3A, monitoring its accretion behavior
in order to characterize accretion flows in the binary environment. Table
\ref{tab:TWA} presents the relevant binary and disk characteristics for TWA
3A.

Binary and higher-order multiple systems are observed as a frequent outcome of
star formation \citep{Raghavanetal2010,Krausetal2011}. Our understanding of
binary population statistics has advanced with large-scale imaging and
spectroscopic surveys, yet the impact binarity has on the star-disk
interaction and planet formation remains poorly understood.  The ubiquity of
binaries, along with the growing number of planets found in, and around,
binary systems \citep{Orosz2012,Kaibetal2013,Mugraueretal2014}, motivates a
detailed characterization of the binary-disk interaction.

Close binaries deviate most from the single-star paradigm where orbital
dynamics are capable of sculpting the distribution and flows of disk
material. For systems with semi-major axes less than $\sim$100 au, orbital
resonances are capable of dynamically clearing a region of disk material
around the binary, opening the possibility for three stable disks: a
circumstellar disk around each star and an encompassing circumbinary disk
\citep{Artymowicz&Lubow1994}. Theory predicts that rather than completely
damming the inflow of material from the circumbinary disk, accretion will
proceed in discrete, periodic streams that form at the inner edge of the
circumbinary disk. These streams cross the cleared gap supplying mass to small
circumstellar disks or accreting directly onto the stars themselves
\citep{Artymowicz&Lubow1996,Gunther&Kley2002,Munoz&Lai2016}.

The frequency of these streams and their impact on the stellar mass accretion
rate are predicted to be highly dependent on the binary orbital parameters.
Focusing on eccentric, equal-mass binaries, similar to TWA 3A, numerical
simulations predict that every apastron passage ($\phi$=0.5) will induce a
stream of circumbinary material that leads to an accretion event near
periastron passage ($\phi$=0;1). These episodes are predicted to increase the
specific accretion rate by up to a factor of 10. This periodic accretion
behavior has been observed in the T Tauri binary DQ Tau
\citep{Mathieuetal1997,Tofflemireetal2017a}.

The accretion streams predicted by the binary-disk interaction are likely
important astrophysical phenomena at a variety of scales. From giant planet
formation spurring streams across disk gaps \citep{Lubow&DAngelo2006} to
accretion onto binary black holes \citep{Bowenetal2017}, the same physical
processes are at play.  Interferometry and adaptive optics techniques are
beginning to spatially resolve such structures in pre-MS systems
\citep{Becketal2012,Casassusetal2013,Yangetal2017}, but they are unable to
describe their temporal characteristics.  Accretion in short-period, pre-MS
binaries offers a unique regime to probe the dynamics of accretion streams.

\begin{deluxetable}{lcc}
\tablewidth{0pt}
\tabletypesize{\footnotesize}
\tablecaption{TWA 3A System Characteristics}
\tablehead{
  \colhead{Parameter} &
  \colhead{Value} &
  \colhead{References}}
\startdata
P (days)                         & $34.87846 \pm 0.00090$ & 1 \\
$\gamma$ (km s$^{-1}$)           & $+10.17 \pm 0.40$      & 1 \\
$e$                              & $0.6280 \pm 0.0060$    & 1 \\
T$_{\rm peri}$ (HJD-2,400,000)    & $52704.554 \pm 0.063$  & 1 \\
$a$ ($R_\odot$)                   & $46.51 \pm 0.49$    & 1 \\
$q\equiv M_2/M_1$                & $0.841 \pm 0.014$      & 1 \\
$M_1$ ($M_\odot$)                 & $0.6027 \pm 0.0207$    & 1 \\
$M_2$ ($M_\odot$)                 & $0.5072 \pm 0.0158$    & 1 \\
Periastron Separation ($R_\odot$)& $17.30 \pm  0.33$    & 1 \\
Apastron Separation ($R_\odot$)  & $75.72 \pm  0.85$    & 1 \\
$i_{\rm disk}$ ($^\circ$)          & $36$                    & 2 \\
Disk $M_{\rm dust}$ ($M_\odot$)    & $7\times10^{-6}$        & 2 \\
$v$sin$i$ (km s$^{-1}$)          & 20                      & 3 \\
d (pc)                           & $30 \pm 3$        & 1,4\tablenotemark{a} \\
$A_V$                            & $0.04 \pm 0.3$          & 5 \\

\enddata 
\tablerefs{$^{1}$Kellogg et al.\ (2017; assuming $i_{\rm binary}$=$i_{\rm
    disk}$=36$^\circ$), $^{2}$\citet{Andrewsetal2010},
  $^{3}$\citet{Torresetal2003}, $^{4}$\citet{Ducourantetal2014}, $^{5}$This
  work}
\tablenotetext{a}{Kinematic distance derived using the
  \citet{Ducourantetal2014} position, proper motion, and convergent point with
  the Kellogg et al.\ (2017) $\gamma$ velocity.}
\label{tab:TWA}
\end{deluxetable}

\section{OBSERVATIONS \& DATA REDUCTION}
\label{obsdr}

In order to characterize the accretion behavior of TWA 3A, we have conducted a
long-term, moderate-cadence, optical photometric monitoring campaign using the
Las Cumbres Observatory (LCO) and SMARTS queue-scheduled facilities. 

\subsection{LCO 1m Network}
\label{lcogt}

The LCO 1m network comprises nine 1m telescopes located across four global
sites: Siding Springs Observatory (Australia), SAAO (South Africa), CTIO
(Chile), and McDonald Observatory (USA).  Spanning $\sim$12 TWA 3A orbital
periods, our observations were made between 2014 May and 2016 April. Observing
visits were scheduled 20 times per orbit while the target was visible (airmass
$<$ 2), corresponding to a cadence of 42 hr. Each visit consisted of three
images in the $UBVR$ filters. All data are reduced by the LCO pipeline
applying bad-pixel, bias, dark, and flat-field corrections. The three images
per filter per visit are than aligned, combined, and fit with astrometric
solutions using standard IRAF tasks.

\subsection{SMARTS 1.3m}
\label{smarts}

The SMARTS 1.3m telescope at CTIO is outfitted with the ANDICAM detector. Our
program requested every-other-night visits of TWA 3A while it was visible
(airmass $<$ 2) between 2014 December and 2016 July. Each visit consisted of
three images in the $B$ and $V$ filters. Data are reduced with the SMARTS
pipeline, which applies bias and flat-field corrections. Each set of images
per visit are aligned, combined, and fit with an astrometric solution using
standard IRAF tasks.

\subsection{Photometry and Calibration}
\label{photcal}

For each telescope (LCO; SMARTS) and filter set ($UBVR$; $BV$), SExtractor
\citep{Bertin&Arnouts1996} is used to perform automated source detection and
photometry on each image producing time-series instrumental magnitudes. A
source catalog for each data set is then created from spatial matching of the
astrometric solutions. Due to poor seeing and/or telescope focus, flux from
TWA 3A and TWA 3B could not be consistently separated. As such, SExtractor
parameters were optimized to photometer the entire TWA 3 system.

Each source catalog is then fed into an ensemble photometry routine following
the \citet{Honeycutt1992} formalism. By selecting non-varying comparison stars
interactively, variations from airmass and nightly observing conditions are
corrected, resulting in relative-magnitude light curves for each star in the
field of view (FOV). We note that this correction does not include color
information, which leaves some small systematic error, especially in $U$-band
where atmospheric corrections are most color dependent.

Relative ensemble magnitudes are then transformed to apparent magnitudes using
non-varying stars in the LCO FOV for which published empirical or derived
photometry exists. Five such stars are present in the LCO FOV (TYC 7213-797-1,
7213-391-1, 7213-1239-1, 7213-933-1, 7213-829-1). Their $V$ magnitudes range
from 11.18 to 11.57 and ($B$-$V$) colors span 0.71 to 2.00. Using empirical
measurements where available, we draw $B$ and $V$ values from the All-Sky
Compiled Catalogue \citep{Kharchenko2001}. For $R$-band, we use a colorless
transformation from the Carlsberg Meridian Catalog 15 (CMC15
\citealt{NBI2014}) $r^\prime$ to $R$ ($R$=$r^\prime-$0.22$\pm$0.12) derived
from 3690 overlapping stars between CMC15 and the Lowell Observatory
Near-Earth Object Search \citep{Skiff2007}. Without empirical $U$-band
measurements available, we use the fitted apparent $U$-band magnitudes derived
by \citet{Pickles&Depagne2010}.  From these five stars we compute zero-point
and color transformations that are applied to the rest of the field.
Systematic errors associated with this calibration procedure are determined
from the root-mean-squared deviation between the transformed published values
of the calibrating stars in color-magnitude space. They are 0.18, 0.22, 0.25,
and 0.07 mag for $U$, $B$, $V$, and $R$ bands, respectively. (Systematic
errors are propagated through the mass accretion rate derivation that follows
and are presented in Figure \ref{fig:mdpg}.)

Not all five of the calibration stars in the LCO FOV are present in the
smaller 6\arcmin \ FOV of the ANDICAM CCD. To transform the SMARTS data to
apparent magnitudes, we bootstrap the apparent magnitudes derived for
non-varying stars in the LCO FOV that overlap with the SMARTS FOV and use
those to determine zero-point and color transformations.

\begin{figure*}[!tbh]
  \centering 
  \includegraphics[keepaspectratio=true,scale=1.0]{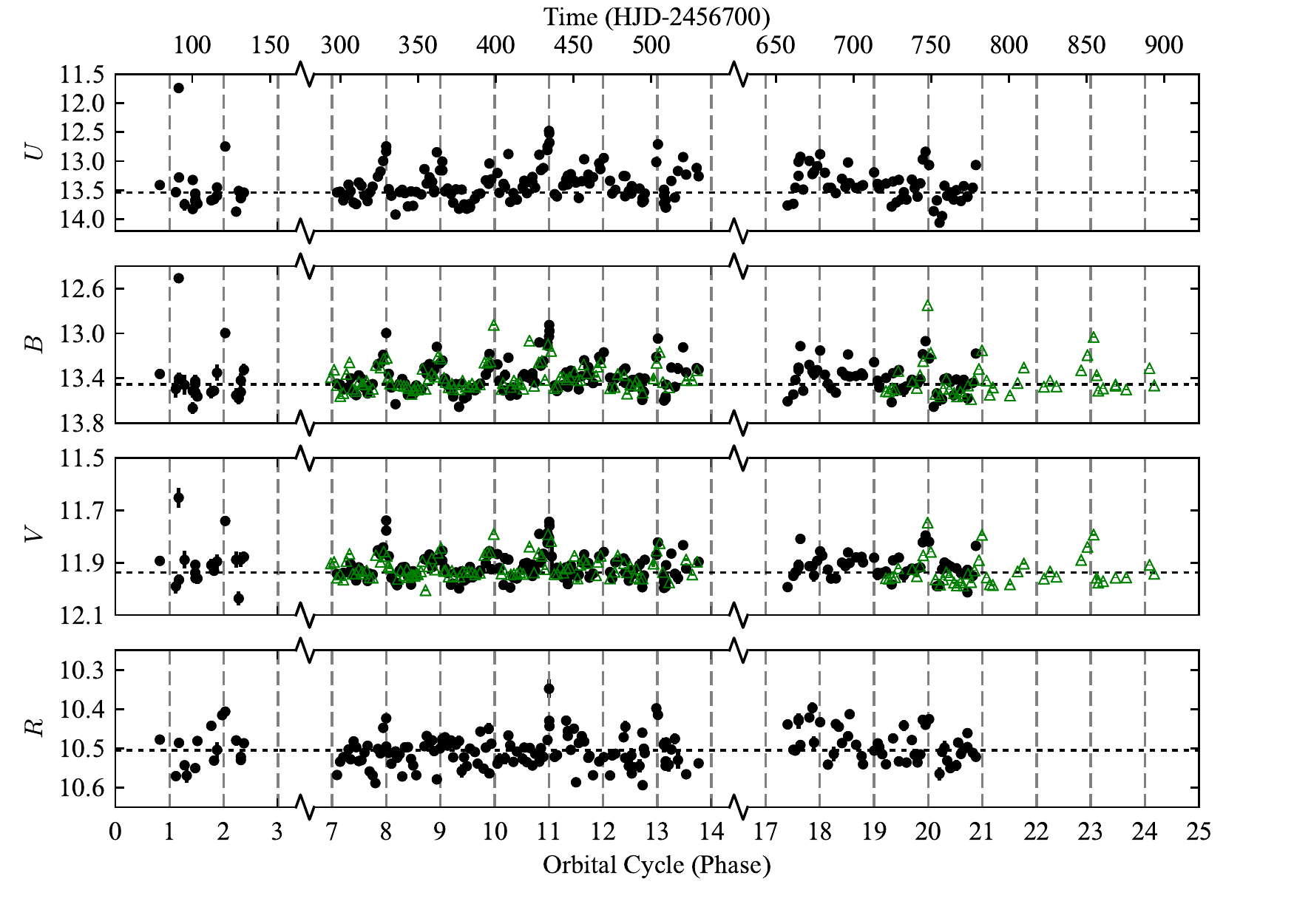}
  \caption{TWA 3 $UBVR$ light curves plotted against arbitrary orbital cycle
    number. The corresponding heliocentric Julian date (HJD) is presented on
    the top y-axis. LCO and SMARTS data are represented as circles and
    triangles, respectively. Vertical dashed lines mark periastron
    passages. Horizontal dashed lines mark the quiescent flux level.}
  \label{fig:lc}
\end{figure*}

\begin{figure}[!tb]
  \centering 
  \includegraphics[keepaspectratio=true,scale=0.5]{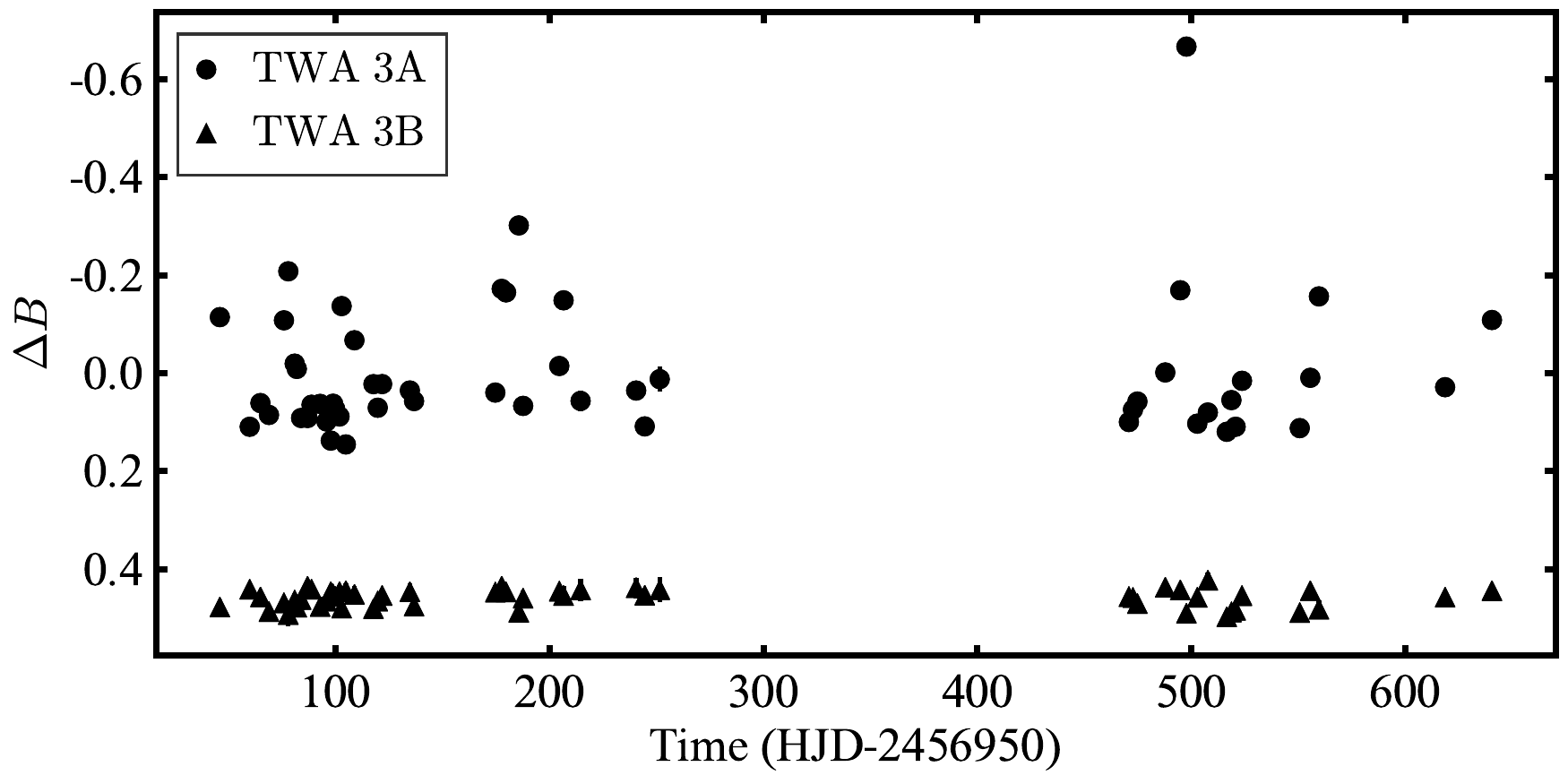}
  \caption{SMARTS $B$-band light curve of TWA 3A (circles) and TWA 3B
    (triangles) from observations in which point-spread-function photometry
    can separate the contribution from both components. The TWA 3A quiescent
    magnitude is $\sim$0.38 mag brighter than TWA 3B.}
  \label{fig:lcpsf}
\end{figure}

\section{Analysis}
\label{analysis}

\subsection{Light Curve Variability}
\label{lc}

Figure \ref{fig:lc} presents the $U$-, $B$-, $V$-, and $R$-band light curves for
TWA 3 plotted against an arbitrary orbital cycle number set to 1 for the first
observed periastron passage. In each panel, vertical dashed lines mark the
TWA 3A periastron passage and horizontal dotted lines mark the quiescent value
(average of orbital phases 0.2 to 0.4). Brightening events near periastron
passages are seen consistently, having the largest increase in $U$-band. These
events very closely match the accretion behavior predicted for eccentric
binaries.

To ensure variability observed in the TWA 3 system is indeed from the
disk-bearing binary and not the tertiary, we perform point-spread-function
photometry on a subset of the SMARTS $B$-band images where the light from each
component can be reliably separated. Figure \ref{fig:lcpsf} displays the
result where TWA 3A is the clear source of variability. The standard deviation
of these light curves are 0.14 and 0.02 mag for TWA 3A and TWA 3B,
respectively. In the following, we assume all variability in the TWA 3 system
results from the spectroscopic binary.

Before assigning all optical variability to changes in the TWA 3A accretion
rate, we inspect our light curves for contributions from stellar flares
(magnetic reconnection events at stellar surfaces). In high-cadence photometry
of DQ Tau, stellar flares with amplitudes greater than $\Delta U$=0.32 mag
were found to have a temporal contribution of $\sim$3\%
\citep{Tofflemireetal2017a}. Assuming the same contribution in TWA 3A
corresponds to six measurements. In moderate-cadence photometry, however,
these events would likely go undetected having similar amplitudes and colors
as accretion variability. Only if a measurement were to contain a large flare
or a flare peak, where the photometric color is typically bluer than
accretion, would it stand out from the underlying accretion variability 
\citep{Kowalskietal2016}.

One measurement in our light curves has a color, magnitude, and temporal
behavior that suggest it contains a stellar flare. Occurring at orbital cycle
$\sim$1.2 in Figure \ref{fig:lc}, it is the brightest $U$-band measurement by
0.7 mag and the bluest in ($U$-$B$) by 0.2 mag, well separated from both
observed distributions. The associated flux would correspond to a factor of
$\sim$15 increase in the mass accretion rate compared to other measurements at
similar orbital phases and a factor of two greater than the next highest
measurement. A measurement only 5 hr later, however, falls securely within the
remaining spread for that orbital phase. Accretion events of this scale are
expected to be rare and to evolve over much longer timescales
(e.g.\ \citealt{Codyetal2014}). And critically, the three $U$-band images
combined in this measurement show a rapid $\sim$0.2 mag decline over $\sim$3
minutes. Given these characteristics, we conclude this measurement contains a
stellar flare and remove it. In the following analysis, we assume the
remaining variability is due to changes in the TWA 3A accretion rate.

\subsection{Mass Accretion Rate}
\label{mdot}

Flux-calibrated $U$-band photometry can be used to derive a mass accretion
rate with knowledge of the distance and extinction to the source, the
photospheric $U$-band flux in the absence of accretion, and the stellar mass
and radius. With values for the distance and stellar parameters, we determine
the extinction and photospheric properties following
\citet{Herczeg&Hillenbrand2014}. First, we fit a spatially resolved Keck LRIS
spectrum of TWA 3A \citep{Herczegetal2009} with a library of empirical
weak-lined T Tauri star (WTTS) spectra. The spectra are fit with three free
parameters: a flux normalization, an additive accretion spectrum, and the
extinction. Our results are consistent with those in
\citet{Herczeg&Hillenbrand2014}, namely, $A_V$=0.04 ($\pm$0.30 mag) and a
combined TWA 3A spectral type of M4.1 ($\pm$0.3 subclasses). Second, the
best-fitting WTTS spectrum is convolved with a $U$-band filter to determine
the underlying photospheric contribution of the binary.

Because our $U$-band measurements are for the entire TWA 3 system, we use a
spatially resolved Keck LRIS spectrum of TWA 3B
\citep{Herczegetal2009} to determine its $U$-band
contribution. Assuming a distance of 30 pc, we extinction correct the $U$-band
measurements and convert them to $U$-band luminosities. Subtracting the
contribution from TWA 3B and the underlying TWA 3A photosphere, we arrive at
the TWA 3A $U$-band accretion luminosity. Using the model-dependent, empirical
relation derived in \citet{Gullbringetal1998}, we calculate the total accretion
luminosity from the $U$-band as follows:
\begin{equation}
  {\rm log}(L_{\rm Acc}/L_\odot) = 
  1.09\ {\rm log}(L_{\rm {\it U_{\rm excess}}}/L_\odot)+0.98.
\label{eqn:U2L}
\end{equation}

For a single star, the mass accretion rate can be determined from the
accretion luminosity with the following:
\begin{equation}
  \dot{M}\simeq \frac{L_{\rm Acc}R_\star}{GM_\star} \left(
  1-\frac{R_\star}{R_{\rm in}} \right)^{-1},
\label{eqn:MA}
\end{equation}
where $R_{\rm in}$ is the magnetospheric disk truncation radius from which
material free falls along magnetic field lines (typically assumed to be
5$R_\star$; e.g.\ \citealt{Johnstoneetal2014}).

Our measurements, however, are of the combined accretion luminosity from two
stars with different masses and radii. Without a theoretical consensus for
which star should predominantly receive the mass from circumbinary accretion
flows, we assume that each star accretes at the same rate. The accretion
luminosity emitted from the primary star alone, $L_{\rm Acc,1}$, becomes
\begin{equation}
L_{\rm Acc,1}=L_{\rm Acc,Total} \left( 1+q\frac{R_{\star,1}}{R_{\star,2}}
\right)^{-1},
\label{eqn:lacc1}
\end{equation}
where $q$ is the mass ratio. The total mass accretion rate for the binary is
then
\begin{equation}
  \dot{M}\simeq 2 \frac{L_{\rm Acc,1}R_{\star,1}}{GM_{\star,1}} \left(
  1-\frac{R_{\star,1}}{R_{\rm in}} \right)^{-1}.
\label{eqn:MA_2}
\end{equation}
For the stellar radii, we use the \citet{Dotteretal2008} stellar evolution
models to compute the average radii for 0.6 and 0.5 $M_\odot$ stars between 5
and 10 Myr. They are 1.06 and 0.99 $R_\odot$, respectively. $R_{\rm in}$ is
set to the canonical single-star value of 5$R_\star$. The derived accretion
rates range between 0.8$\times$10$^{-11}$ and 2.4$\times$10$^{-10}$ $M_\odot$
yr$^{-1}$, in good agreement with previous measurements
\citep{Muzerolleetal2000,Herczegetal2009}. Since the stars have near equal
masses, the choice to split the accretion rate equally between the two
corresponds to only a $\pm$$\sim$5\% difference from assigning all the accretion
to one star.

\begin{figure}[!t]
  \centering
  \includegraphics[keepaspectratio=true,scale=0.5]{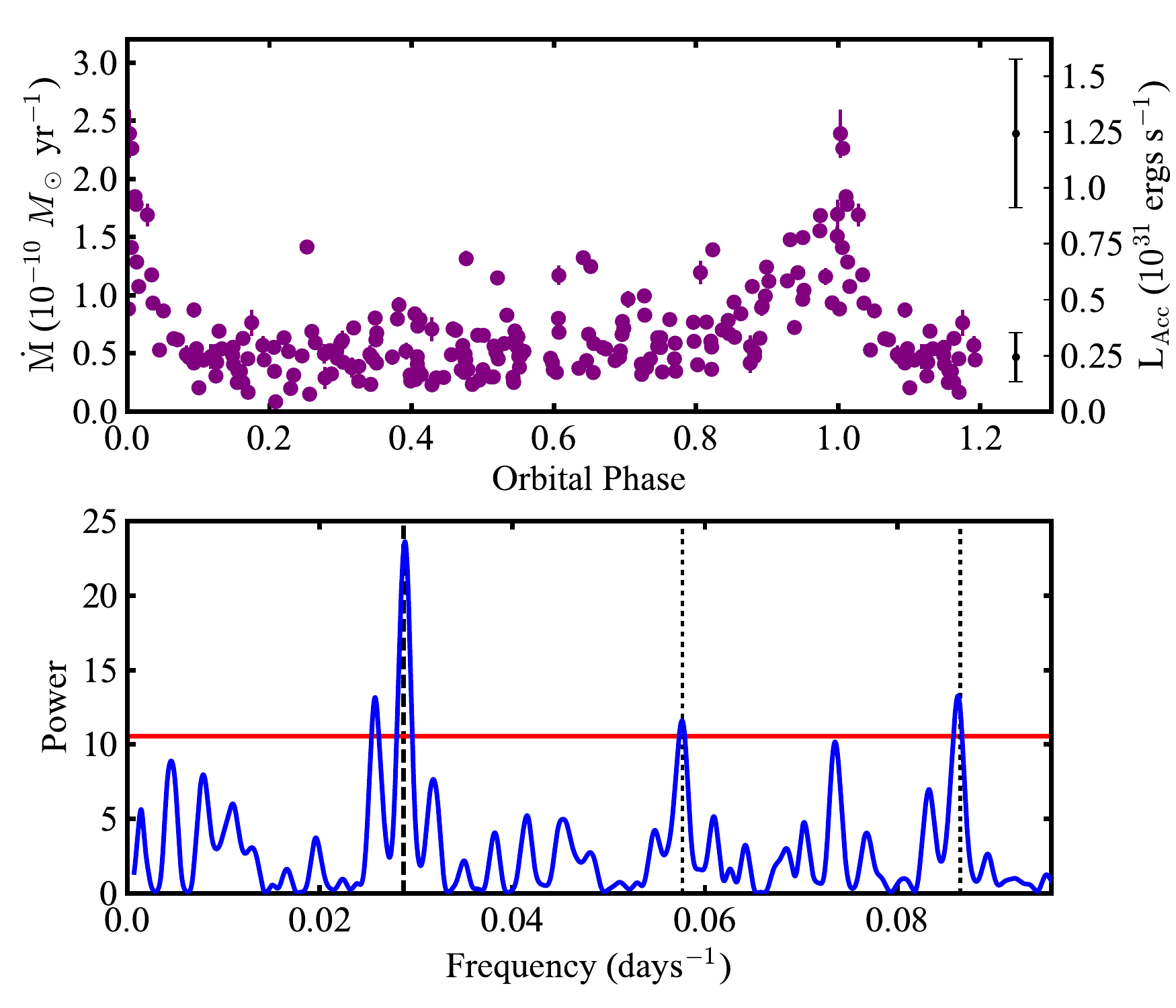}
  \caption{{\bf Top:} TWA 3A mass accretion rate and accretion luminosity
    phase-folded about the orbital period.  Black error bars on the right
    correspond to the propagation of the systematic error of our photometric
    calibration. Because the systematic error is relative, we present it for
    the quiescent and peak accretion rate values. {\bf Bottom:} Periodogram of
    the mass accretion rate measurements.  Significant power is found above
    the 99\% false-alarm probability (horizontal red line) consistent with the
    orbital period (vertical dashed line).}
  \label{fig:mdpg}
\end{figure}

The top panel of Figure \ref{fig:mdpg} presents the mass accretion rate
phase-folded about the orbital period. The repeated enhanced accretion events
observed near periastron increase the accretion rate by a factor of
$\sim$3--10 from the quiescent value.

\subsection{Accretion Periodicity}
\label{period}

To determine the significance of the periodic accretion behavior, we perform a
Lomb-Scargle periodogram \citep{Scargle1982} on the accretion rate
measurements. The bottom panel of Figure \ref{fig:mdpg} presents the power
spectrum. A significant peak is observed above the 99\% false-alarm
probability with a period of 34.67$\pm$0.14 days, in good agreement
(1.5$\sigma$) with the binary orbital period. (Error in the accretion rate
period is derived from a $10^6$ iteration Monte Carlo bootstrap simulation
using random sampling with replacement of the $\dot{M}$ and HJD measurement
pairs \citealt{Pressetal1992}.)  Additional higher-frequency peaks occurring
at two and three times the peak frequency result from the varying and
non-sinusoidal morphology of the enhanced accretion events. The small peak to
the left of the primary does not remain significant after filtering the data
at the primary frequency.

\begin{figure}[!t]
  \centering
  \includegraphics[keepaspectratio=true,scale=0.5]{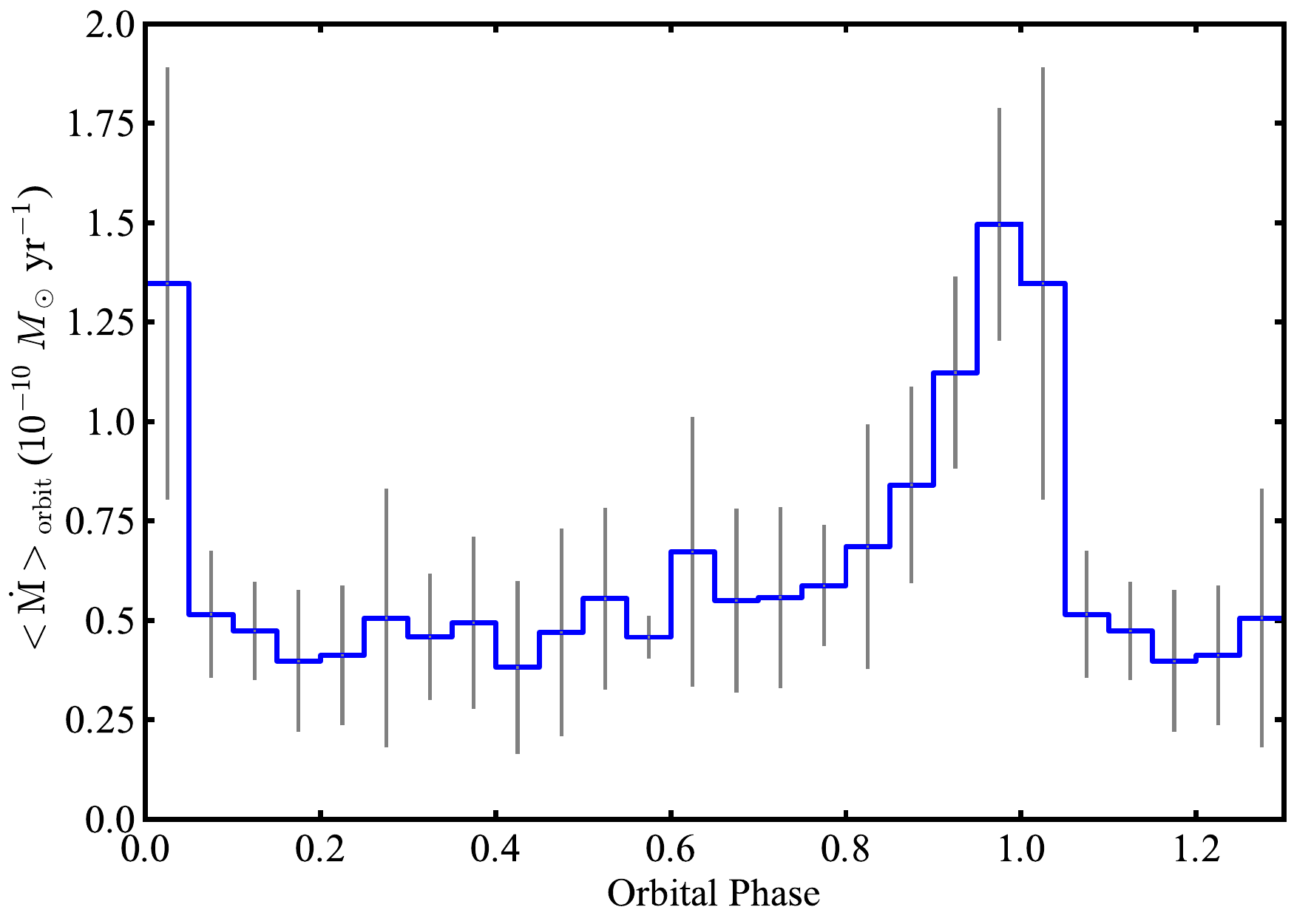}
  \includegraphics[keepaspectratio=true,scale=0.5]{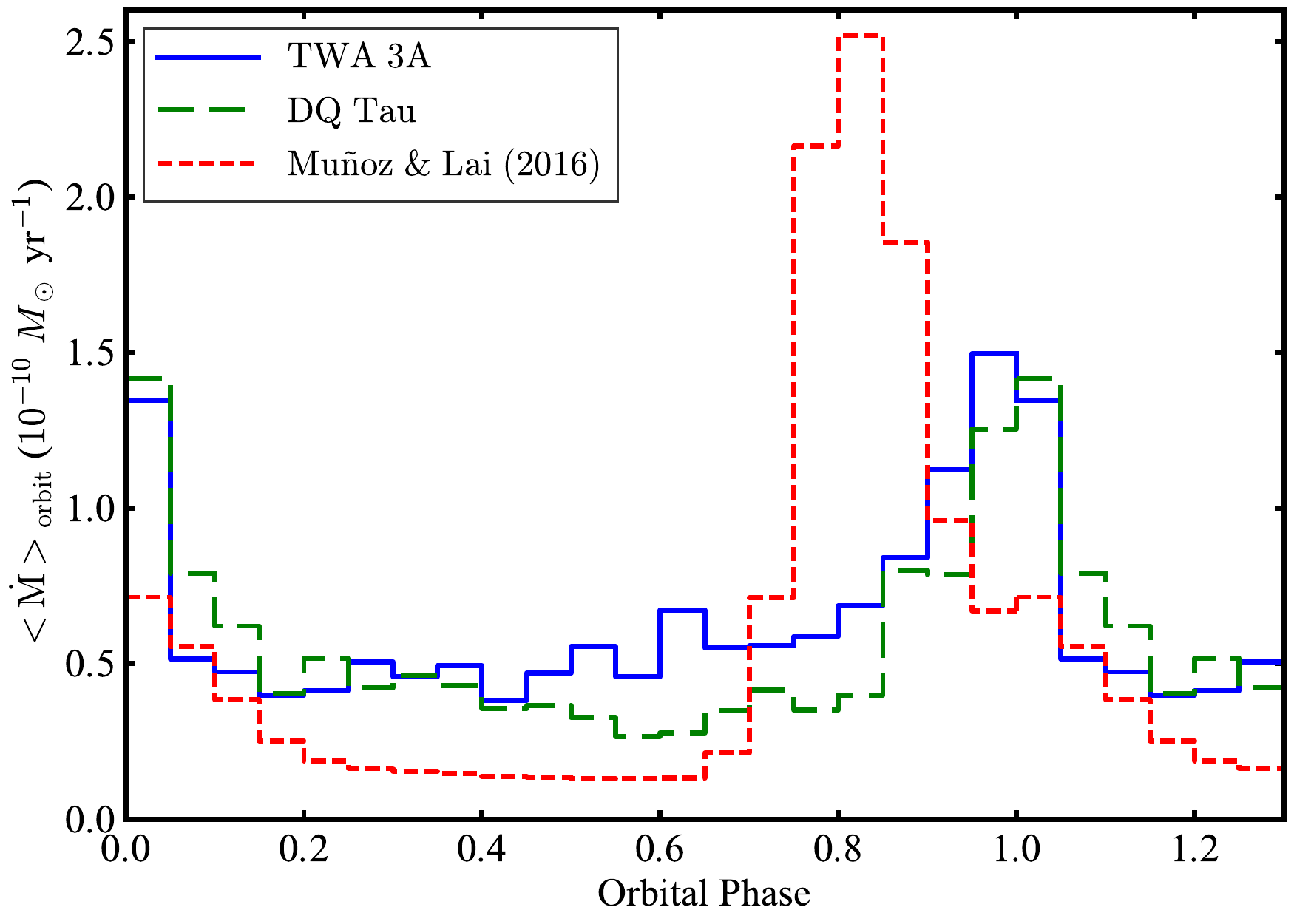}
  \caption{{\bf Top:} TWA 3A median accretion rate as a function of orbital
    phase. Error bars represent the standard deviation within each phase bin.
    {\bf Bottom:} Comparison of the TWA 3A accretion rate profile with DQ Tau
    \citep{Tofflemireetal2017a} and a numerical simulation from
    \citet{Munoz&Lai2016} for a binary with similar orbital parameters.}
  \label{fig:PAA}
\end{figure}

\subsection{Accretion Rate Profile}
\label{profile}

The morphology of enhanced accretion events contains information on the
interaction between the binary orbit and the circumbinary mass flows. In order
to compare our observations with numerical simulations, we create an average
accretion rate profile as a function of orbital phase. Breaking the
orbital-phase-folded data into bins of $\phi$=0.05 (our sampling rate), we
calculate the median mass accretion rate for each bin and set the bin error as
its standard deviation. The result is presented in the top panel of Figure
\ref{fig:PAA} where, on average, the accretion rate is elevated between
orbital phases 0.85 and 1.05, reaching a peak near periastron of $\sim$4 times
the average quiescent value.

In the bottom panel of Figure \ref{fig:PAA} we compare the TWA 3A accretion
rate profile to a simulation of binary accretion \citep{Munoz&Lai2016} and to
the DQ Tau accretion rate profile \citep{Tofflemireetal2017a}. Both have been
normalized to the TWA 3A average mass accretion rate. The
\citet{Munoz&Lai2016} simulation shown is a 2D hydrodynamical model using the
adaptive-mesh-refinement code AREPO \citep{Springel2010} for an equal-mass
binary with an eccentricity of 0.5. Ten consecutive orbital periods of the
simulation were used to create the accretion profile.

\section{Discussion}
\label{disc}

Figures \ref{fig:mdpg} and \ref{fig:PAA} provide conclusive evidence that
accretion in TWA 3A is strongly influenced by the binary orbit, leading to
periodic accretion bursts near periastron passage. This behavior is largely
consistent with the prediction of numerical simulations supporting the
scenario that periodic streams of mass are capable of carrying material across
a cleared gap to the central binary. Similar behavior has only been this
clearly observed in one other binary, DQ Tau. The binary UZ Tau E
\citep{Jensenetal2007} and the protostar LRLL 54361 \citep[unknown
  period]{Muzerolleetal2013} are also intriguing sources that have shown hints
of phase-dependent accretion.

Direct comparisons to numerical simulations can begin to constrain the
dynamics of accretion streams. Despite a notable phase offset, there is good
agreement between TWA 3A and the \citet{Munoz&Lai2016} model. Making more
in-depth comparisons, however, is not straightforward.  The difficulty lies in
the relevant hydrodynamic and magnetic scales that effect stable circumstellar
disk material in short-period systems. 

In high-resolution hydrodynamic simulations, each star develops a stable
circumstellar disk that collects and organizes incoming circumbinary stream
material. Accretion events in this case result from a combination of tidal
torques that each star induces on its companion's disk and the interaction of
the circumbinary streams with the circumstellar disks. Without magnetic
fields, material is accreted once it reaches the stellar surfaces. These
stable circumstellar disks have the effect of regularizing accretion events
from orbit-to-orbit.

In short-period systems, however, there is a close match between the outer
dynamical truncation radii of circumstellar disks imposed by binary orbital
resonances and the inner magnetic truncation radii typically assumed for
single stars. If magnetic fields are capable of disrupting or reducing the
size of stable circumstellar disks, there may be a direct interaction of
stream material with the stellar magnetosphere that is not captured by current
models.

We note that in lower-resolution simulations by \citet{Gunther&Kley2002}, each
star develops a marginally resolved circumstellar disk, yet their accretion
rate profile peaks directly at periastron, matching the observations of TWA 3A
and DQ Tau. It is unclear whether resolved circumstellar disks or different
disk treatments (viscosity, radiative cooling, viscous heating, etc.) are
responsible for these differences in the timing and amplitude of accretion
events. Short of including magnetic fields in three dimensions, a study
varying the disk properties and inner accretion radii would provide a more
suitable comparison with the data presented here.

Finally, the TWA 3A and DQ Tau average accretion rate profiles are striking
similar. Although they have similar orbital parameters, DQ Tau exhibits much
more variability from orbit to orbit (compare Figure \ref{fig:mdpg} to Figure
7 in \citealt{Tofflemireetal2017a}). Yet, on average, their profiles are very
similar in shape and amplitude. Regularity in TWA 3A accretion events when
compared to DQ Tau, may be the result of a lower overall accretion rate
($\sim$10$^{-2}$), or perhaps its larger semi-major axis permits some amount
of stable circumstellar disk material that regularizes accretion events.

\section{Summary}
\label{sum}

With a long-term, densely sampled, optical photometric monitoring campaign, we
have characterized the accretion behavior of the young binary TWA 3A. Here, we
summarize the main results of our work:

\begin{enumerate}

\item Spatially resolved photometry reveals that accretion variability from
  the spectroscopic binary, TWA 3A, is the dominant source of optical
  variability in the combined light of the TWA 3 system.

\item From $U$-band observations we derive the TWA 3A mass accretion rate as a
  function of time. Periodic accretion events are observed near each
  periastron passage. On average, the accretion rate is elevated between
  orbital phases 0.85 to 1.05, reaching a peak of $\sim$4 times the quiescent
  value. 

\item The observed behavior is in good agreement with numerical simulations,
  providing strong evidence for periodic circumbinary accretion streams in TWA
  3A. This is only the second clear case of orbital-phase-dependent accretion
  in a pre-MS binary.

\item These are some of the first data that can begin to constrain the
  dynamical properties of accretion flows. Comparisons with current models are
  limited, however, as they do not include the disruptive effect magnetic
  fields have on stable circumstellar disks.

\item The TWA 3A average accretion rate profile is remarkably similar to that
  of DQ Tau (a shorter-period system with similar mass ratio and eccentricity;
  \citealt{Tofflemireetal2017a}) despite DQ Tau's larger variability from
  orbit to orbit.

\end{enumerate}

\acknowledgments 

The authors would like to thank Lisa Prato for providing the TWA 3A
spectroscopic orbital solution prior to its publication.  This work makes use
of observations from the LCO network and the CTIO 1.3 m telescope operated by
the SMARTS Consortium. This research has made use of the VizieR catalogue
access tool, CDS, Strasbourg, France. B.~T.\ acknowledges funding from Sigma
Xi Honors Society and from the University of Wisconsin-Madison Graduate
School.

\end{document}